\documentclass{kapproc}
\usepackage{procps}
\usepackage{epsf}

\kluwerbib
\newcommand{\etal}{{et al}\/.}

\begin{document}
 
\articletitle{The environments of radio-loud quasars}

%\articlesubtitle{This is an Article Subtitle}
\author{J.M. Barr,~~~ M.N. Bremer}
\affil{Physics Department, Bristol University, UK}
\email{j.barr@bristol.ac.uk, }

\author{J.C. Baker}
\affil{Department of Astronomy, Berkeley, USA}

\begin{abstract}   
We have obtained multi-colour imaging of a representative,
statistically complete sample of low-frequency selected
($S_{408\mathrm{MHz}}>0.95$Jy) radio loud quasars at intermediate ($0.6<z<1.1$)
redshifts. These sources are found in a variety of environments, from
the field through to rich clusters. We show that statistical measures
of environmental richness, based upon single-band observations are
inadequate at these redshifts for a variety of reasons. Environmental   
richness seems correlated with the size and morphology of the radio
source, as expected if the energy density in the radio lobes is
approximately the equipartition value and the lobes are in pressure
equilbrium with a surrounding intragroup/cluster medium. Selecting on
radio size therefore efficiently selects dense galactic sytems at
these redshifts.

\end{abstract}

\section{Introduction}

Our sample of quasars were a subset of the statistically complete
Molonglo Quasar Sample (MQS) (Kapahi \etal, 1998)\cite{kapahi98}, with
redshifts 0.65$<$z$<$1.10. and with $0<RA<14$. Several sources were
randomly excluded due to observing-time limitations. Radio
luminosities of the sources ranged between $27.4<\mathrm{Log}(P_{408\mathrm{MHz}})<28.2$,
a factor 5 lower than the most luminous (3CR) sources at comparable
redshifts.

Most fields were imaged in the optical using either EFOSC or EFOSC2 on
the ESO 3.6m telescope with a field of view of 5'x 5', with some
observed with the CTIO and AAT 4m telescopes. Filter bands were chosen
to straddle the 4000\AA \ break giving the greatest contrast for early
type galaxies at the appropriate redshift (typical images being complete to 
$R\sim 24$). A subset of these fields
were imaged in $J$ and $K$ using IRAC2 on the ESO/MPI 2.2m. with a   
smaller (2'x 2') field-of-view (typically complete to $K\sim 18.5$).

\section{Single filter clustering statistics}

\begin{figure}  
\begin{center}
\epsfysize 6cm
\epsfbox{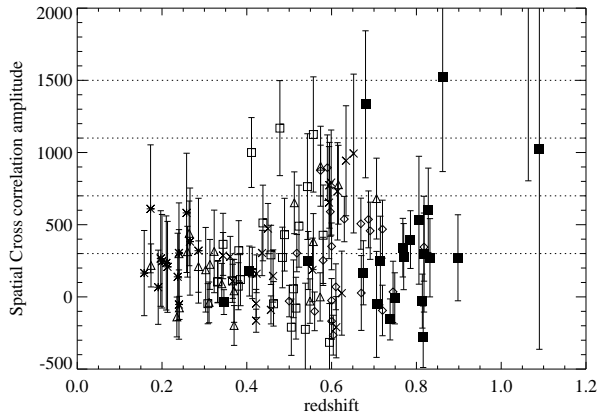}
\end{center}
\caption{Values of the spatial cross correlation amplitude as a function of redshift for Radio loud quasar fields. Filled squares, this work. Diamonds, Wold \etal (2000). Triangles, Yee \& Ellingson (1993). Squares, Ellingson \etal  (1991). Crosses, Yee \& Green (1987). Asterisks, McClure \& Dunlop (2000). The horizontal lines represent Abell classes 0,1,2,3.}
\label{bgq}
\end{figure}

We analysed clustering of faint galaxies around the quasars in two
ways. Firstly, we carried out ``traditional'' statistical tests of
clustering on single-band data (or at least not colour-selecting the
galaxies), as practiced by others ({\it e.g.} see Wold \etal , these
proceedings).  The spatial cross-correlation amplitude B$_{gq}$ and the
magnitude-limited overdensity within 0.5~Mpc, N$_{0.5}$ (Hill \&
Lilly, 1991) \cite{hill91} were computed for our fields in either the  
$R$ or $I$ band. The results for B$_{gq}$ are shown in Fig.\ \ref{bgq}.
They show broad agreement with those of Wold \etal \ (2000) 
\cite{wold00} and McClure \& Dunlop, (2000) \cite{mcclure00} to a
redshift of 0.9. Above this, our data are not deep enough to detect a
significant portion of cluster galaxies. Normalising to the expected
cluster luminosity function then causes spurious results with large   
errors. Our observations support the notion that RLQs inhabit a wide
variety of environments as found in other studies ({\it e.g.} Yee \&
Green, 1987) \cite{yee87}.

Additionally we find several cases where B$_{gq}$ is misleading. This
may be for several reasons. Firstly, interpretation of the the
statistical tests in terms of environmental richness assumes that the
quasar is at the centre of any (roughly circular) overdensity. At high
redshift, clusters are generally not relaxed spheroidal systems, and
may have several bright ellipticals within them, rather than one
central dominant galaxy. We find that our quasars are not always at   
the centres of overdensities, and in some cases there are other
galaxies that have colours and magnitudes of first-ranked cluster 
ellipticals within the overdensity.

Secondly, at the magnitude levels of interest, the number density of
background objects, and the significant variation in that quantity on
the scale-lengths of distant clusters make it very difficult to
determine the background in one band to subtract from any overdensity.
Thirdly, even when an obvious overdensity is detected, it can
sometimes be made up of galaxies with the wrong optical-IR colours to
be at the redshift of the quasar. In at least two cases we find a
overdensity which gives a large value of B$_{gq}$ but is a foreground
agglomeration, found to be so by reference to optical and infrared
colours. Detections such as these can lead to the misidentification of
clusters and an increase in the average value of the cross-correlation
amplitude.

\section{Multicolour analysis}

Because of the problems with single-band statistical measures of   
clustering, we used the colours of galaxies in the quasar fields to
estimate clustering. Specifically, we determine the colours expected
for passively evolving ellipticals at the centres of clusters at the
redshift of each quasar and compare these to the colours of the
observed galaxies.

Optical $V-I,R-I$ colours can be used to isolate the 4000\AA \ break
for ellipticals at $z>0.6$.  Near-infrared $J$ and $K$ filters are also
useful for identifying galaxies above a redshift of 0.5. Specifically,
objects with a $J-K$ colour of $\sim$ 1.7 and above are galaxies at
redshift 0.5 and higher (Pozzetti \& Manucci, 2000).
\cite{pozzetti00} Spirals at these redshifts can also be isolated due
to their red $J-K$ (due to dust reddening), but relatively blue optical 
colours. The $J-K$ colour sets unresolved high redshift galaxies
completely apart from stars and brown dwarfs which have $J-K<1.5$.

\begin{figure}
\begin{center}
\epsfysize 2cm
\epsfbox{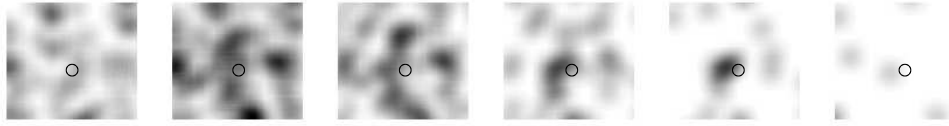}
\end{center}
\caption{The density of $21<I<23$ objects in the 5'x5' field of MRC B0450-221 
($z=0.9$). The figure shows progressively redder objects in 0.5
Mag bins from $0.5<V-I<1.0$ on the left through to $3.0<V-I<3.5$ on
the right. The position of the quasar is indicated by the circle; it is on 
the edge of an overdensity of $2.5<V-I<3.0$ objects.}
\label{0450}
\end{figure}

For each object we determine the distribution of a magnitude limited
sample of objects in the field as a function of their colour. Figure \ref{0450}
shows an example of this.  Each greyscale square is the smoothed
surface density of objects with $21<I<23$ with colours ranging between
$0.5<V-I<3.5$ in bins of 0.5 mag in colour in the 5'x 5' field of a  
$z=0.9$ quasar. The grey levels are normalised to give the same object
density for the same grey level in each colour bin. Most objects are
bluer than $V-I<2$. An overdensity of red objects close to the quasar 
can be seen in the $2.5<V-I<3$ bin, the colour expected for ellipticals
at $z=0.9$. $J$ and $K$ imaging of the central region confirm that these
red objects have $J-K$ and $I-K$ colours of ellipticals at
z=0.9. Narrow band imaging in redshifted [OII] and follow-up
spectroscopy confirm faint star forming galaxies in the field at the
quasar's redshift, but not in the region of the red overdensity. This 
is a clear case of a cluster around a z=0.9 quasar and is discussed in
detail in Baker \etal \ (2001) \cite{baker01}.

\begin{figure}
\begin{center}
\epsfysize 2cm
\epsfbox{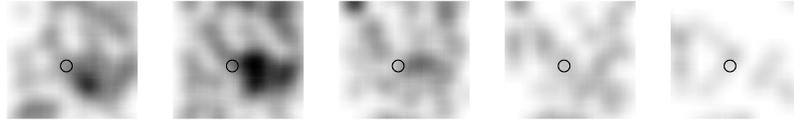}
\end{center}
\caption{The density of objects in the field of MRC B0346-279. The  
figure shows the density of objects with $0<R-Z'<2.5$ in 0.5 Mag
bins. It can clearly be seen that the large value of N$_{0.5}$ is   
caused by clustering of blue ($0.5<R-Z'<1.0$) objects to the West of the
quasar.}
\label{0346}
\end{figure}

In contrast Figure \ref{0346} shows the distribution of faint objects around a
$z=1$ quasar. The R-band image clearly shows a strong overdensity of
$22<R<24$ close to the quasar, leading to a strong N$_{0.5}$ signal,
indicating at first sight that the quasar inhabits a rich cluster .
However, the $R-Z'$ colour distribution peaks at $0.5<R-Z'<1.0$, far
too blue for ellipticals at $z=1$. Similarly the $R-K$ colours of
these objects are also too blue to be ellipticals at the quasars
redshift. Thus we have a case where the colour information rules out
this overdensity being a cluster at the quasar redshift (unless the
morphological mix is extremely unusual).

\begin{figure}
\begin{center}
\hbox{
~~~~~~~~~~~~~~~~~~~\epsfysize 3cm
\epsfbox{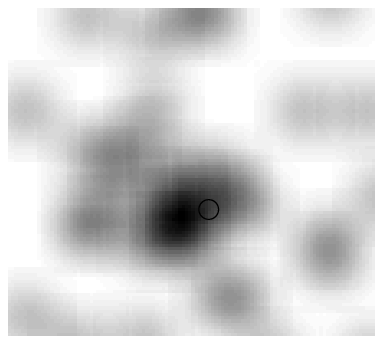}
~~~~~\epsfysize 3cm
\epsfbox{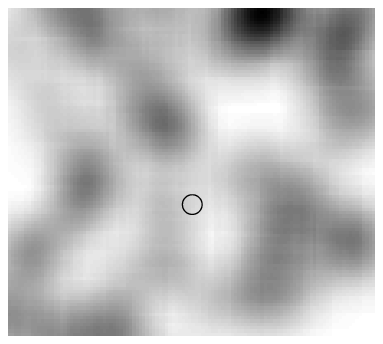}
}
\end{center}
\vspace{-7mm}
\caption{Distribution of the density of $20<I<22$ objects with $1.4<R-I<1.6$ for two 5'x5' fields centred on MRC quasars at z=0.77. Both greyscales are normalised to the same surface densities.}
\label{1224_1247}
\end{figure}

Figure \ref{1224_1247} compares the spatial distribution of objects   
with $20<I<22$ and $1.4<R-I<1.6$ in the fields of two quasars with
almost identical redshifts ($z=0.77$). The first cluster shows a clear
overdensity of objects with this colour, consistent with the colour
ellipticals at $z=0.77$, the second does not (and shows no obvious   
clustering of objects of any other colour). The $R-K$ colours of the
objects around the first quasar are again consistent with being in a
cluster associated with the quasar.

Similar analysis of other objects in the sample lead to detections of
several other systems of galaxies, from compact groups to clusters
along with other systems that at first sight appear to be groups or
clusters at the quasar redshift, but have colours consistent with
lower (or different) redshift objects.

\section{A possible correlation with radio size}

The range of radio luminosities for our sample covers less than an
order of magnitude, and the lookback time difference between the
lowest and highest redshift members is less than 2Gyr, so we cannot
probe if these quantities correlate with environmental richness.
However, we may expect a correlation with the extent of the radio
lobes in low-frequency selected radio sources.  If the radio lobes are
in pressure equilibrium with a surrounding medium, then if their
internal energy densities/pressures are close to (or scale roughly as)
the equipartition value, there should be relationship between true
(unprojected) angular size and the pressure of the external medium
({\it e.g.} see Miley 1980 \cite{miley80}, Bremer \etal, 1992)
\cite{bremer92}.  For sources that evolve self similarly, $P_{equip}
\propto (\frac{F}{\theta^3})^{4/7}$ where $F$ is the flux of the
source (essentially constant in a flux-limited sample) and $\theta$ is
the angular size of the source. The smaller the source, the higher the
equipartition pressure, and therefore the higher the external pressure
(assuming source pressures scale with the equipartition value, at
least statistically). Sources surrounded by a dense ICM should
therefore be smaller than those in the field, or lower mass groups.

Excluding flat spectrum point sources, there are seven sources which
have a radio extent of less than 20''. We find four obvious clusters,
of which three are confirmed by IR data. The fourth has an obvious
overdensity with the correct optical colours, but requires IR
data to confirm the nature of the constituent galaxies.

Of the remaining nine larger objects, only three show signs of
overdensities with the right colours. Two of these are asymmetric
sources which appear to have groups associated with their disturbed
radio morphology, {\it e.g.}  Bremer \etal \ (2001) \cite{bremer01}.

\section{Summary}

%\begin{enumerate}

%\noindent \item 
Powerful low-frequency selected radio-loud quasars at redshifts
$0.6<z< 1.1$ exist in a wide variety of environments, from the field
through compact groups to rich clusters.

%\noindent \item 

The quasar is not always directly centred on any overdensity we
find, nor is any overdensity confined to within 0.5 Mpc of the active
galaxy. This has the effect of making clustering statistics like
B$_{gq}$ and N$_{0.5}$ rather blunt tools for analysing individual
clusters. We find that colours must be used when trying to distinguish
clusters from their backgrounds. In particular near infrared colours
can be used to accurately extract the high redshift galaxy population.

%\noindent \item 
The smallest extended sources are more likely to be classified as
being in clustered environments (though not necessarily in
clusters). This is to be expected from equipartition
arguments. Factoring in an estimate of true source size and excluding
core-dominated sources results in an extremely efficient way of
selecting fields containing high redshift groups and clusters. The
richness of these systems vary from field to field, but (given the
results of Martin Hardcastle in these proceedings) we could expect
comparable richnesses to many X-ray selected clusters at similar
redshifts. Blind, wide-field optical searches for clusters find
systems of comparable richness (but with far lower
efficiency). Despite the obvious effectiveness of colour selection in
detecting these systems, follow-up 8m imaging and spectroscopy is
still required to determine the parameters of the systems.

%\end{enumerate}


\begin{chapthebibliography}{1}
\bibitem[]{baker95}Baker J.C., Hunstead R.W., Brinkmann W., 1995, MNRAS, 277, 553
\bibitem[]{baker01}Baker J.C., Hunstead R.W., Bremer M.N., Bland-Hawthorn J., 
Athreya R. M. Barr J.M., 2001, AJ, 121, 1821.
\bibitem[]{bremer92}Bremer M.N., Crawford C.S., Fabian A.C., Johnstone R.M., 1992, MNRAS, 254, 614
\bibitem[]{bremer01}Bremer, M.N., Baker, J.C., Lehnert, M., 2001, MNRAS submitted.
\bibitem[]{ellingson91}Ellingson  E., Yee H.K.C., Green R.F., 1991, ApJ, 371, 49
\bibitem[]{hill91}Hill G.J., Lilly S.J., 1991, ApJ, 367, 1
\bibitem[]{kapahi98}Kapahi V.K., Athreya R.M., Subrahmanya C.R., Baker J.C., 
Hunstead R.W., McCarthy P.J., van Breugel W., 1998, ApJS, 118, 327
\bibitem[]{mcclure00}McClure R.J., Dunlop J.S., 2000, MNRAS, submitted 
(astro-ph/0007219)
\bibitem[]{miley80}Miley G., 1980, ARAA, 18, 165
\bibitem[]{pozzetti00}Pozzetti L., Mannucci  F., 2000, MNRAS, 317, L17
\bibitem[]{wold00}Wold M., Lacy M., Lilje P.B., Serjeant S., 2000, MNRAS 316, 
267
\bibitem[]{yee87}Yee H.K.C., Green R.F., 1987, ApJ, 319, 28
\bibitem[]{yee93}Yee H.K.C., Ellingson  E., 1993, ApJ, 411, 43

\end{chapthebibliography}
\end{document}